\documentclass[draft]{agujournal2019}
\usepackage{url} 
\usepackage{lineno}
\usepackage[inline]{trackchanges} 
\usepackage{soul}
\usepackage{amsmath}
\usepackage{graphicx}
\usepackage{makecell}

\draftfalse

\journalname{JGR: Earth Surface}

\begin{document}

\title{First Direct Observations of Internal Flow Structures in a Powder
Snow Avalanche: Turbulence, Instability and Particle
Distribution}

%
%




\authors{I. Calic\affil{1,2}, F. Coletti\affil{2}, B. Sovilla\affil{1}}

\affiliation{1}{Avalanche Formation and Dynamics, WSL Institute for Snow and Avalanche Research SLF, Davos Dorf, Switzerland}
\affiliation{2}{Institute of Fluid Dynamics, Mechanical and Process Engineering, ETH Zurich, Zurich, Switzerland}




\correspondingauthor{Ivan Calic}{ivan.calic@slf.ch}



\begin{keypoints}
\item First direct observations of snow particle motion inside airborne layers of a natural powder snow avalanche using high-speed imaging.
\item Identified large-scale fluctuations exceeding turbulent integral scales, linked to shear instabilities through linear stability analysis.
\end{keypoints}

\begin{abstract}
Powder snow avalanches are highly dynamic, multiphase gravity-driven flows typically composed of a dense basal layer overlain by airborne layers in which snow particles are suspended within a turbulent air phase. Despite extensive work on related systems such as pyroclastic density currents and turbidity currents, all gravity current communities face a fundamental limitation: the lack of direct, high-resolution particle-scale field data. Here, we present the first direct optical observations of individual particle motion inside the airborne layers of a natural powder snow avalanche using high-speed imaging. The flow is segmented into three regions: an initial short living surge, a highly dynamic suspension phase, and a final wake. Across these phases, we quantify flow velocity and turbulence characteristics, including integral length scales, and use image intensity as a proxy for particle concentration. We identify fluctuations exceeding the turbulent integral scale and use linear stability analysis to link them to Kelvin–Helmholtz-type shear instabilities. Observed changes in clustering behavior, stratification, and the decoupling of particles from the flow mark the transition from an unstable suspension layer characterized by a high level of turbulent activity to a stable one dominated by passive snow settling. Together, these findings provide the first empirical constraints on turbulence and instability dynamics in airborne avalanche layers, with direct implications for the refinement of numerical avalanche models and closure schemes in multiphase gravity current simulations.

\end{abstract}

\section*{Plain Language Summary}
Powder snow avalanches are fast-moving flows of snow and air that can travel long distances and pose serious hazards in mountain regions. These avalanches consist of a dense bottom layer and lighter airborne layers where snow particles are suspended in turbulent air. Until now, direct measurements of particle-scale dynamics in the airborne layers have been missing, limiting understanding of how turbulence, mixing, and instabilities develop and preventing accurate model predictions.
In this study, we used high-speed cameras to capture the motion of snow particles inside a natural avalanche. We identified three distinct flow stages: an initial short-lived surge, a highly dynamic suspension phase, and a calmer wake where particles settle. By analyzing particle brightness and motion, we estimated particle concentration, flow velocity, and turbulence characteristics. We observed patterns consistent with shear instabilities, similar to Kelvin–Helmholtz waves, which drive mixing and clustering within the flow.
These first direct measurements of turbulence and particle behavior in airborne avalanche layers provide new insight into how avalanches evolve. The understanding gained from these observations can help refine numerical models, improve predictions of avalanche flow and avalanche hazard, and advance knowledge of turbulent multiphase flows in natural gravity-driven systems.

\section{Introduction}

Gravity currents, including turbidity currents (TCs), pyroclastic density currents (PDCs) and powder snow avalanches (PSAs), are among the most destructive natural flows, posing severe risks to infrastructure and human life \cite{simpson_gravity_1999, meiburg_turbidity_2010, dufek_fluid_2016, lacaze_gravity_2024}. Despite their destructive potential, the internal dynamics that govern their evolution remain incompletely understood. Even under steady boundary conditions, gravity currents exhibit pronounced unsteadiness, like velocity fluctuations, internal surges, and intermittent layering \cite{wells_turbulence_2021, nasr-azadani_mixing_2018, dufek_fluid_2016, breard_inside_2017, kohler_dynamics_2016, sovilla_intermittency_2018}. These unsteady flow features can largely exceed model predictions and complicate hazard assessment and mitigation planning \cite{brosch_destructiveness_2021}. 

For all particle-laden gravity currents, a fundamental limitation is the scarcity of direct, high-resolution field measurements at the particle scale. For TCs, in situ access is limited by their subaqueous environment \cite{parker_experiments_1987, paull_powerful_2018} and for PDCs, by extreme temperature \cite{dufek_fluid_2016, lube_multiphase_2020}. In contrast, PSAs provide a uniquely accessible natural laboratory. They are highly dynamic, multiphase flows with analogies to their volcanic and subaqueous counterparts, yet can be studied at dedicated test sites under controlled release conditions \cite{norem_measurement_1985, ammann_new_1999}. At the same time, PSAs are among the most complex of gravity currents, characterized by high velocities, strong particle–air coupling, and a layered internal structure. A widely accepted conceptual model distinguishes three layers: a dense basal layer, an intermediate transition layer, and an upper suspension layer \cite{sovilla_structure_2015, issler_inferences_2020}. Whether all three layers form, and how they evolve, depends strongly on the physical properties of the snow as it is released and entrained along the avalanche path \cite{kohler_geodar_2018}.

Systematic full-scale measurements of avalanches began in the 1980s with impact-pressure sensors, which provided the first quantitative records of forces and allowed velocity estimates \cite{schaerer_seismic_1980, shimizu_study_1980, mcclung_characteristics_1985, kawada_experimental_1989}. The establishment of permanent test sites, such as  Vallée de la Sionne (VdlS, Switzerland) and Ryggfonn (Norway), enabled repeated, instrumented experiments and expanded observations to internal velocities and flow depths, especially in the avalanche dense layer \cite{norem_measurement_1985, ammann_new_1999, sovilla_measurements_2008, kern_measured_2009, gauer_full-scale_2007, gauer_four_2016}. Ultrasonic anemometers and photography captured air-flow and cloud structures \cite{nishimura_structures_1993, nishimura_measurements_1995, nishimura_velocity_1997}, while more recently radar and seismic methods have captured surging behavior at large scales \cite{kohler_dynamics_2016, sovilla_intermittency_2018}. 
At VdlS, dedicated high-frequency pressure sensors have also been deployed in the powder-snow cloud, allowing avalanche front speed, size, and location to be inferred from air-pressure fluctuations using a dipole approximation \cite{mcelwaine_air_2005}.
Together, these approaches have greatly advanced our understanding of avalanche dynamics, but key processes within the airborne layers remain poorly constrained.

Turbulence is thought to govern the very processes that sustain these airborne layers, entrainment, particle suspension, and clustering \cite{fukushima_numerical_1990, meiburg_turbidity_2010, sovilla_intermittency_2018}. Capturing these processes requires highly resolved velocity measurements. Existing in-situ records remain too coarse to resolve particle-scale dynamics. Available ultrasonic and optical measurements provide only limited constraints on velocity fluctuations and turbulent length scales \cite{nishimura_structures_1993, nishimura_measurements_1995, nishimura_velocity_1997, kern_measured_2009, ito_measurement_2017}.

Only a few measurements of snow-particle sizes exist, reporting values from millimeters to centimeters in the transition layer \cite{schaer_particle_2001}, and mean diameters of 0.2 mm in the suspension layer \cite{rastello_size_2011, ito_measurement_2017}. Because particle size critically controls suspension, settling, and clustering in multiphase turbulence \cite{parker_experiments_1987, breard_inside_2017, lube_multiphase_2020, issler_inferences_2020}, numerical models must account for these particle-size effects. In practice, this is often achieved by turbulence closures (e.g., Reynolds-Averaged Navier-Stokes (RANS) models \cite{fukushima_numerical_1990}) or bulk parameterizations (e.g., depth-averaged models; \cite{vicari_mot-psa_2025}), though these are rarely calibrated by empirical data.

In addition to turbulence, shear-driven instabilities, such as Kelvin–Helmholtz instabilities are hypothesized to influence local mixing and large-scale flow variability \cite{balachandar_turbulent_2010, kostaschuk_causes_2018}. Observations of mesoscale structures and pulses in PSAs suggest that such instabilities may exist \cite{sovilla_intermittency_2018}, but direct, time-resolved measurements confirming their presence and the conditions under which they form are lacking.

Despite advances in experimental methods, modeling, and conceptual understanding, several key aspects of PSA dynamics remain unresolved. These include (i) direct validation of the three-layer model, particularly for the airborne transition and suspension layers, (ii) detailed turbulence characteristics such as velocity fluctuations, particle distributions, and turbulent length and time scales, and (iii) the existence and role of shear-driven instabilities in controlling entrainment, layering, and intermittent surging.
In this study, we address these unresolved gaps by presenting the first direct, high-resolution, in-situ imaging experiments within the airborne layers of a full-scale PSA, complementing radar \cite{ash_two-dimensional_2014} and optical sensor measurements \cite{kern_measured_2009}. Using a high-speed camera array deployed at VdlS, we extract time-resolved particle velocities and flow structures across the entire avalanche. We segment the flow into three regions: early surge, main suspension, and stabilized wake, and analyze each in terms of vertical velocity profiles, turbulence metrics, particle distribution, and potential instability signatures. This dataset provides the first opportunity to quantify particle-scale turbulence and investigate the presence of layered structures and shear-driven instabilities within full-scale PSAs. 

\section{Methods}
To reconstruct the vertical structure and flow dynamics of the avalanche, we employed three complementary measurement technologies: GEODAR (GEOphysical flow dynamics using pulsed Doppler radAR) \cite{ash_two-dimensional_2014}, optical sensors \cite{kern_measured_2009}, and high-speed cameras, that together cover different spatial extents, material concentrations, and vertical positions within the flow. Since GEODAR and optical sensors are well-established and have been extensively described in previous studies, we provide only a brief overview of these systems and focus primarily on the novel high-speed camera measurements and their processing. Data acquisition across all sensors and technologies is automatically triggered by seismic sensors upon avalanche detection, ensuring fully synchronized recordings. Upon trigger, the optical sensors and the high-speed cameras record for 240~s. The GEODAR records for 540~s. 

\subsection{The experimental site Vallée de la Sionne }
\begin{figure}[ht!]
    \centering
    \includegraphics[width=1.0\linewidth]{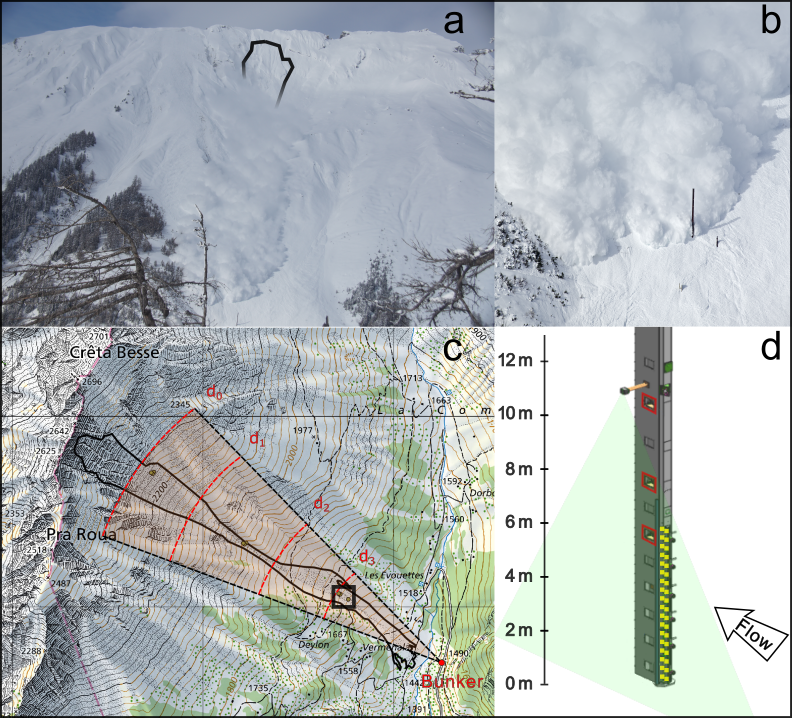}
    \caption{a) Artificially released PSA at VdlS from winter 2015, shown for illustration only. The approximate release area of the studied avalanche \#~20243024 is highlighted. b) A PSA reaches the pylon (Photo: PhotoTerre et Nature O. Born). c) Overview of the VdlS experimental site. The black line outlines the boundaries of the avalanche analyzed in this study. The red shaded area represents the region observed by the GEODAR system. Red dashed lines within this area indicate reference positions ($d_i$) used to relate GEODAR measurements to flow velocity and terrain location (see Fig. \ref{fig:avaVelocity}). The red dot marks the location of the bunker housing the GEODAR. The original map is obtained from the Swiss Geoportal (2025) in Swiss coordinate system CH1903 / LV03 (EPSG:21781). (d) A 20~m high steel pylon located in the runout zone (black rectangle in panel c) is equipped with three high-speed cameras (red boxes) mounted inside protective hatches. A LED panel on the pylon projects a light sheet (green area) perpendicular to the cameras' field of view. Optical sensors (yellow marks) at the base of the pylon measure flow velocity.}
    \label{fig:mapVdlS}
\end{figure}

The study was conducted at the VdlS experimental site, a full-scale avalanche test facility located north of Sion in the Canton of Valais, Switzerland. This site enables controlled studies of avalanches at natural scale (Fig. \ref{fig:mapVdlS}a,b). The avalanche analyzed in this study naturally released between the peaks of Pra Roua and Crêta Besse, on steep slopes ranging from 35° to 45° at an elevation of about 2400~m~a.s.l. (black contour in  Fig. \ref{fig:mapVdlS}c). From the release area, the avalanche descended roughly 800 m in elevation before reaching the instrumented obstacles in the runout zone at 1650~m~a.s.l (black rectangle in  Fig.~\ref{fig:mapVdlS}c). The distance from the release area to the pylon was 2100 m, with the slope gradually decreasing to 25° near the measurement location. The avalanche finally stopped at an elevation of 1450~m after traveling a total distance of 2400 m. 
A central component of the measurement setup is the GEODAR, a frequency-modulated  continuous wave radar installed on a reinforced bunker located near the avalanche runout zone (red dot in Fig. \ref{fig:mapVdlS}c). It observes the avalanche from release to deposition, primarily capturing the dense basal layer, while its 57 mm wavelength penetrates the dilute airborne layers with little attenuation, thereby providing a broad overview of the avalanche’s macroscopic behavior \cite{ash_two-dimensional_2014, kohler_dynamics_2016}. Moving target identification (MTI) plots, space-time diagrams of radar reflectivity, are subsequently generated from the GEODAR data, allowing identification of different flow regimes along the avalanche path, including fronts, surges, and transitions between flow layers \cite{kohler_geodar_2018}.

At a horizontal distance of approximately 600~m from the radar, a 20~m tall steel pylon, instrumented with various sensors (Fig.~\ref{fig:mapVdlS}d), was erected within the runout zone (black rectangle in Fig.~\ref{fig:mapVdlS}c). Velocity measurements were obtained within the lower 6 m of the pylon using optical sensors \cite{kern_measured_2009}. These paired sensors, spaced 12.5~cm apart, provide horizontal velocities by cross-correlating their signals in regions of sufficiently high particle concentration. However, because they require strong signal contrast, their applicability in the more dilute suspension layers is limited.

\subsection{High-Speed Camera Setup}
To extend the measurements into the airborne layers, a new high-speed camera array was deployed for the first time in this work, allowing particle-scale dynamics to be captured across the transition and suspension layers. When combined with optical sensor data, this setup enables to extend the vertical reconstruction of flow velocity from the dense basal layer into the airborne suspension.
Three high-speed cameras (Photonfocus DR4-D1952-L01-GT, CMOS sensor) were installed laterally inside sealed pylon hatches at heights of 10.5~m (high), 7.5~m (mid) and 5.5~m (low) above ground level (Fig.~\ref{fig:mapVdlS}d). Each camera covered approximately 1~m in height (±0.5~m around its mounting position), corresponding to observation windows of 11–10~m, 8–7~m, and 6–5~m above ground level, respectively, with gaps of 1–2~m between the observed regions. The hatches protect the cameras from impacts, snow intrusion, and harsh winter conditions, with protective glass fitted in openings for imaging. For optimal particle visualization, a high-intensity LED (light-emitting diode) panel was mounted externally on the pylon. The light beam was shaped by Fresnel lenses into a 12°~x~60° planar lightsheet passing perpendicularly through the cameras' field of view (FOV) at a distance of approximately 1~m from the pylon (green area in Fig.~\ref{fig:mapVdlS}d). This configuration ensures that only particles within this narrow optical plane are illuminated, allowing precise conversion of pixels into physical units. 
Cameras recorded at full HD resolution (1920~x~1080~px) with 1000~FPS (frames per second), using Zeiss Dimension 12~mm f/2 lenses. Each 4-minute recording produced up to 2~TB of data. Optical distortions at the image edges were minimal, as analysis was restricted to the central region of each camera’s FOV, as described below.

\subsection{Image-Based Extraction of Velocity, Concentration and Particle Size}
\label{sec:imagBased}
\begin{figure}[ht!]
    \centering
    \includegraphics[width=1.0\linewidth]{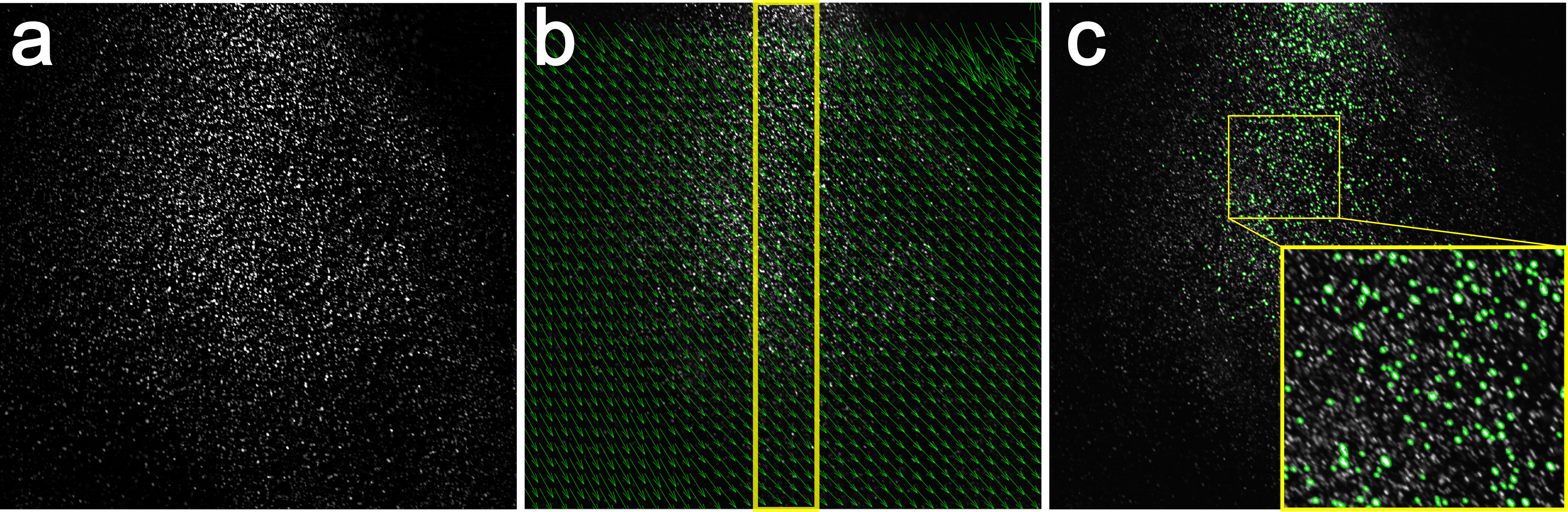}
    \caption{Image processing workflow for velocity and particle size extraction. (a) Preprocessed image after background subtraction and contrast enhancement, used as input for further analysis. (b) Resulting velocity field obtained by PIV. Green arrows indicate particle displacements converted to velocities; the yellow box marks the vertical band used for subsequent high-resolution temporal analysis. (c) Detected particles with sizes estimated from fitted ellipses (green). The yellow rectangle shows a zoomed region with particle identification.} 
    \label{fig:method}
\end{figure}

To extract quantitative flow parameters from the high-speed video recordings, we applied a sequence of image processing steps, including Particle Image Velocimetry (PIV), particle concentration estimation, and particle sizing (Fig. \ref{fig:method}). 
PIV was used to compute spatially and temporally resolved velocity fields from consecutive image pairs. All images were preprocessed by subtracting a static background and applying local contrast enhancement to isolate the illuminated particles (Fig. \ref{fig:method}a). Each image was divided into interrogation windows of 32~×~32 px, and particle displacements were calculated using the PIVlab software of \citeA{stamhuis_pivlab_2014}  (Fig. \ref{fig:method}b). For high-resolution temporal analysis, a vertical band from the center of each velocity field was extracted (yellow box in Fig. \ref{fig:method}b) and horizontally averaged, producing a one-dimensional vertical velocity profile at each time step. This approach yields a space-time velocity map for each camera, providing continuous representation of the velocity evolution (Fig. \ref{fig:PIVraw}). The PIV results were compared and found quantitatively consistent with particle tracking velocimetry (PTV) analysis of the same data, using an in-house algorithm extensively employed to track natural snowfalls \cite{nemes_snowflakes_2017, li_settling_2021}. PIV processing was favored as it is more robust in conditions of high particle concentration \cite{adrian_particle_2011}.

A relative measure of particle concentration was derived from the preprocessed images by computing the ratio of illuminated (bright) pixels to the total number of pixels in the FOV, with bright pixels identified using adaptive thresholding based on local intensity (Fig. \ref{fig:method}a). This image concentration serves as a proxy for local particle concentration rather than an absolute value, since it is influenced by illumination conditions, particle size, and potential particle overlap, but allows comparative analysis across time and height.
Individual particles were then identified from binarized images, and their sizes were estimated from the minor axis of best-fit ellipses (Fig. \ref{fig:method}c). The inset in Fig. \ref{fig:method}c highlights a zoomed-in region to illustrate particle detection. Diffraction effects are negligible at this scale given the optical setup parameters \cite{adrian_particle_2011}. These include a pixel size of 10~$\mu$m, a focal length of 12~mm, and an aperture of f/2. Under these conditions, the sizes of the bright imaged objects extending over multiple pixels reliably represent the resolved snow particles. Particles smaller than the pixel size (about 0.8~mm) are not resolved. The size distributions presented in section \ref{sec:flowRegTurPar}, however, peak at diameters above 2~mm, indicating that particles too small to be resolved did not represent the majority of the suspended mass in the imaged regions. The exception is represented by the recording of the early surge by the lower camera. As will be shown in section \ref{sec:flowRegions}, this faced high concentration of fast moving snow, complicating the detection and segmentation of individual particles.

\section{The Avalanche \# 20243024}
\subsection{Event Overview and Flow Evolution}
Avalanche \#~20243024 occurred spontaneously and was detected by the monitoring system at 05:19:39 AM on 2~December 2023. This instant is taken as t = 0 for all time series, with t representing the elapsed time since detection. 

\begin{figure}[ht!]
    \centering
    \includegraphics[width=0.9\columnwidth]{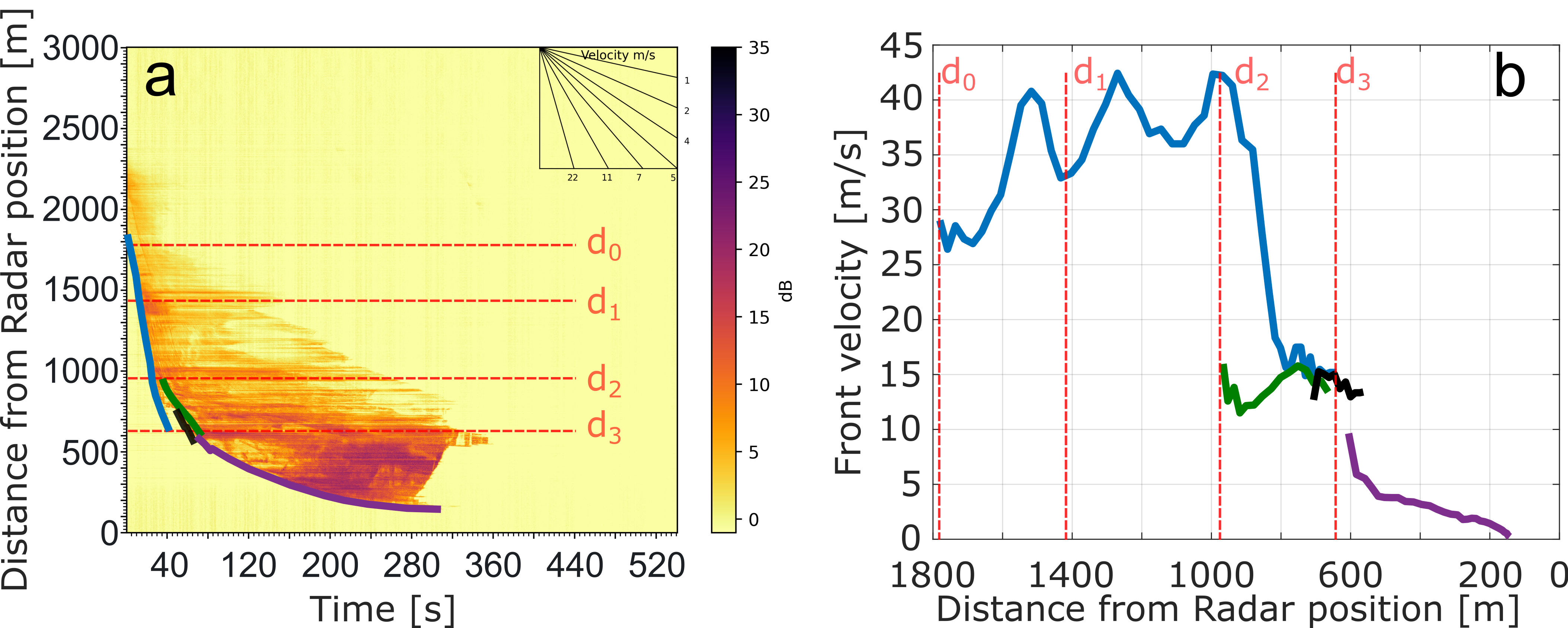}
    \caption{(a) MTI plot from the GEODAR of avalanche \#20243024. (b) Front velocities extracted from the MTI plot. Dashed red lines denote the positions in the avalanche path as shown in Fig. \ref{fig:method}. Colored lines distinguish the separate surges in both the MTI and front velocity plot.}
\label{fig:avaVelocity}
\end{figure}

The avalanche released in dry, cold snow and initially developed as a PSA. Air temperature increased downslope from –8 °C in the release area to –3 °C in the runout zone, conditions that would normally support a sustained PSA from release to deposition. However, as the flow descended, it entrained progressively warmer and wetter snow from deeper layers, altering its properties and driving a transition toward a warmer, denser avalanche \cite{kohler_geodar_2018}. By the time it reached the runout zone and the pylon instrumentation, the avalanche was in a decaying phase. Front velocities (Fig. \ref{fig:avaVelocity}b) from the GEODAR MTI plot (Fig. \ref{fig:avaVelocity}a) confirm that it had entered a deceleration stage. 

The MTI plot (Fig. \ref{fig:avaVelocity}a) shows that the avalanche initially advanced as a single front. At the exit of the couloir at d$_2$ (1000~m in the MTI plot), one of the key reference positions along the path (Fig. \ref{fig:mapVdlS}),  localized surges developed, some detaching from the main body. These surges are thought to originate from internal instabilities or roll-wave activity at the surface of the dense basal layer \cite{kohler_dynamics_2016}. 

The avalanche entered the radar range at d$_0$ with a velocity of 25 m s$^{-1}$, then accelerated toward the couloir entry (d$_1$). Between d$_1$ and d$_2$, the front velocity oscillated between 35 and 45 m s$^{-1}$, before undergoing a sharp deceleration starting from d$_2$. Notably, the deceleration phase coincided with the appearance of the surges. Upon reaching the pylon (d$_3$), the front velocity had decreased to about 15 m s$^{-1}$, with only minor variations between individual surges (Fig. \ref{fig:avaVelocity}b).
 
As the flow reached the runout zone, the supply of dry, cold material available for suspension diminished. This lack of additional mass intake into the airborne layers prevented the evolution of a distinct transition layer, leaving the suspension dominated by air entrainment and particle settling. The absence of a well-defined transition layer therefore provides a unique opportunity to investigate the settling, dilution, and stratification dynamics of the suspension without interference from strong intermittency or large surges. 

\subsection{Data at the Pylon Location}

\begin{figure}[ht!]
    \centering
    \includegraphics[width=0.9\linewidth]{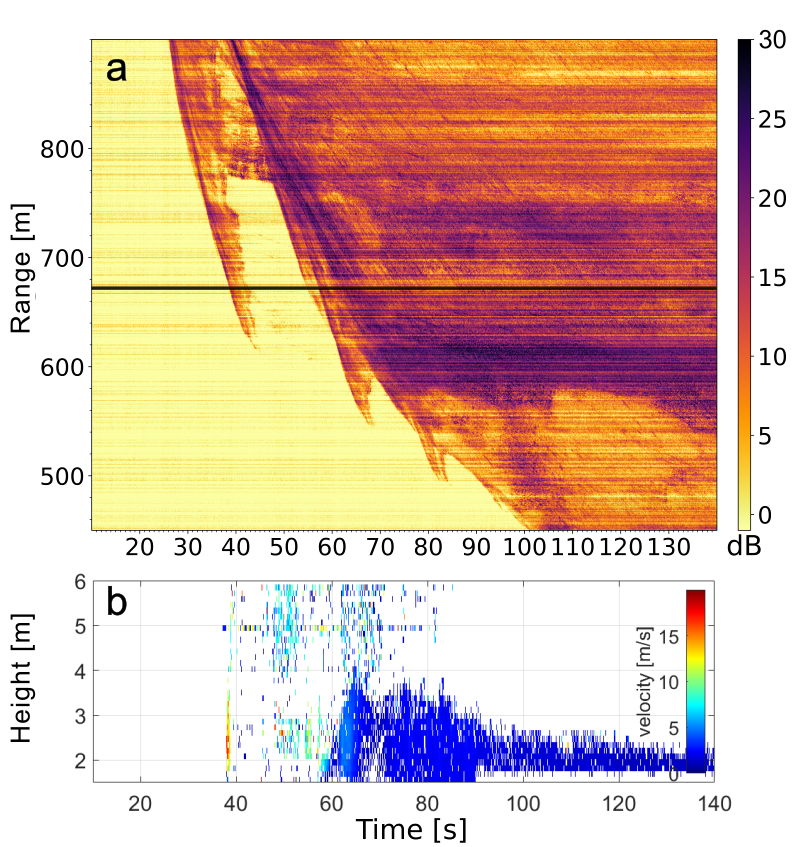}
    \caption{a) Zoomed section of the GEODAR derived MTI plot focused around the position of the pylon (horizontal black line). b) Corresponding velocities measured by the optical sensors on the pylon. The avalanche slid on an old snow cover up to 1.5~m thick.}
    \label{fig:geodarOptical}
\end{figure}
The avalanche was recorded by the GEODAR, optical sensors, and high-speed cameras \cite{calic_vallee_2025}. Figure \ref{fig:geodarOptical} presents a combined view of the GEODAR MTI plot and the velocity data from the optical sensors. The MTI plot in Fig. \ref{fig:geodarOptical}a is a zoomed-in section of the data shown in Fig. \ref{fig:avaVelocity}a, providing a more detailed view of the avalanche dynamics near the pylon. It highlights two main flow phases: an initial fast, short-lived detached surge followed by the main avalanche body, a slower, more sustained dense layer with superimposed smaller surges.

The optical sensors (Fig. \ref{fig:geodarOptical}b) also captured both flow phases. They continued to register velocity signals, even during periods where the GEODAR no longer detected coherent movement. This indicates that particle concentrations remained sufficient for detection. The resulting height–time velocity maps also reveal a well-defined, slower basal layer (dark blue colors in Fig. \ref{fig:geodarOptical}b) overlaid by a faster-moving suspended particle phase, with a sharp boundary separation.

\begin{figure}[ht!]
    \centering
    \includegraphics[width=1\linewidth]{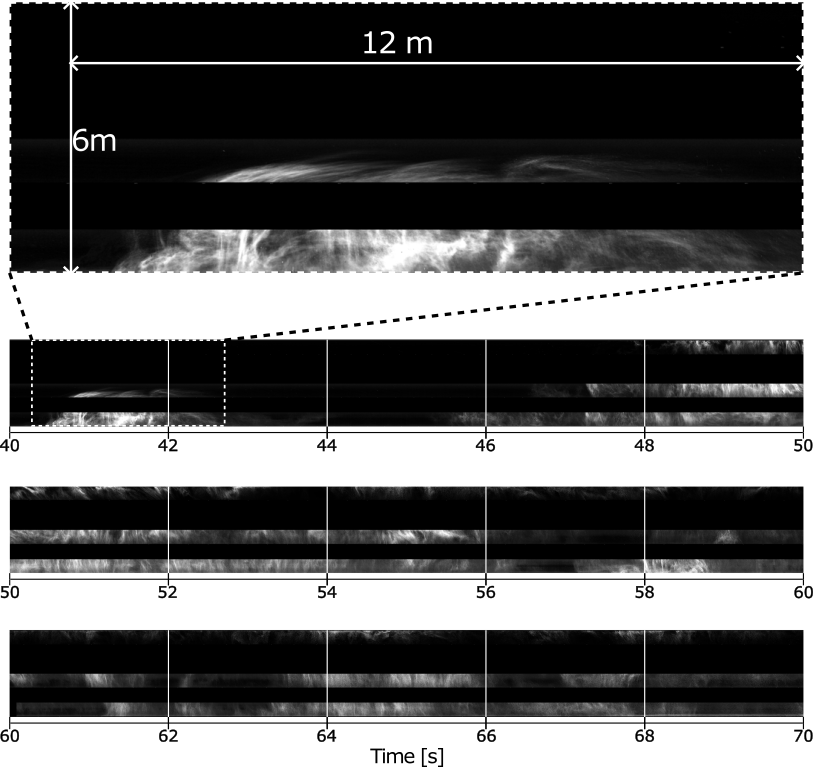}
    \caption{Space-time visualization of the avalanche flow recorded by the high-speed camera array. Each panel shows a stitched sequence of central vertical strips (1~m in height) from three cameras, creating a panoramic view from 5~to~11~m above ground. The upper image provides a zoomed-in view of the initial flow, highlighting the early surge. Black horizontal spacers indicate gaps between the cameras’ fields of view, corresponding to vertical regions not captured by the cameras. A high-resolution version of this figure is provided in the auxiliary material, allowing detailed inspection by zooming.}
    \label{fig:rawStitched}
\end{figure}
The high-speed cameras provided continuous recordings of the avalanche’s airborne layers. Figure \ref{fig:rawStitched} presents a panoramic view of the avalanche evolution over time, generated from central vertical strips (10~px~wide) extracted from 30,000 consecutive frames, covering the first 30~s of the event. The gaps in between the cameras are highlighted by black horizontal spacers.

The upper zoomed-in panel in Fig. \ref{fig:rawStitched} highlights the arrival of the fast, fully detached surge, lasting for approximately 2~s and a total length of approximately 12~m. During this phase, the total flow height remains mostly below the mid-positioned camera at 7.5~m, with no material visible in the uppermost strip. The upper interface of the surge exhibits wave-like structures that may be indicative of shear-driven instabilities at the interface between the suspension and the ambient air, as observed in TCs and other particle-laden density flows \cite{meiburg_turbidity_2010, lacaze_gravity_2024}. These features at the upper interface are indicative of shear-driven instabilities that may promote mixing between the fast-moving surge and the ambient air.

The surge is followed by a gap, during which no visible material appears in front of the cameras, consistent with optical sensor data (Fig. \ref{fig:geodarOptical}). Shortly afterward, the main body of the flow arrives, exhibiting a markedly different structure from the initial surge. Material becomes sequentially visible, first in the lowest, then in the mid, and finally in the highest camera’s FOV. During this phase, the avalanche depth progressively increases, eventually exceeding the height of the uppermost camera.

\section{Results}
\subsection{High-Speed Imaging Measurements}
\label{sec:highSpeed}
\begin{figure}[ht!]
    \centering
    \includegraphics[width=0.8\linewidth]{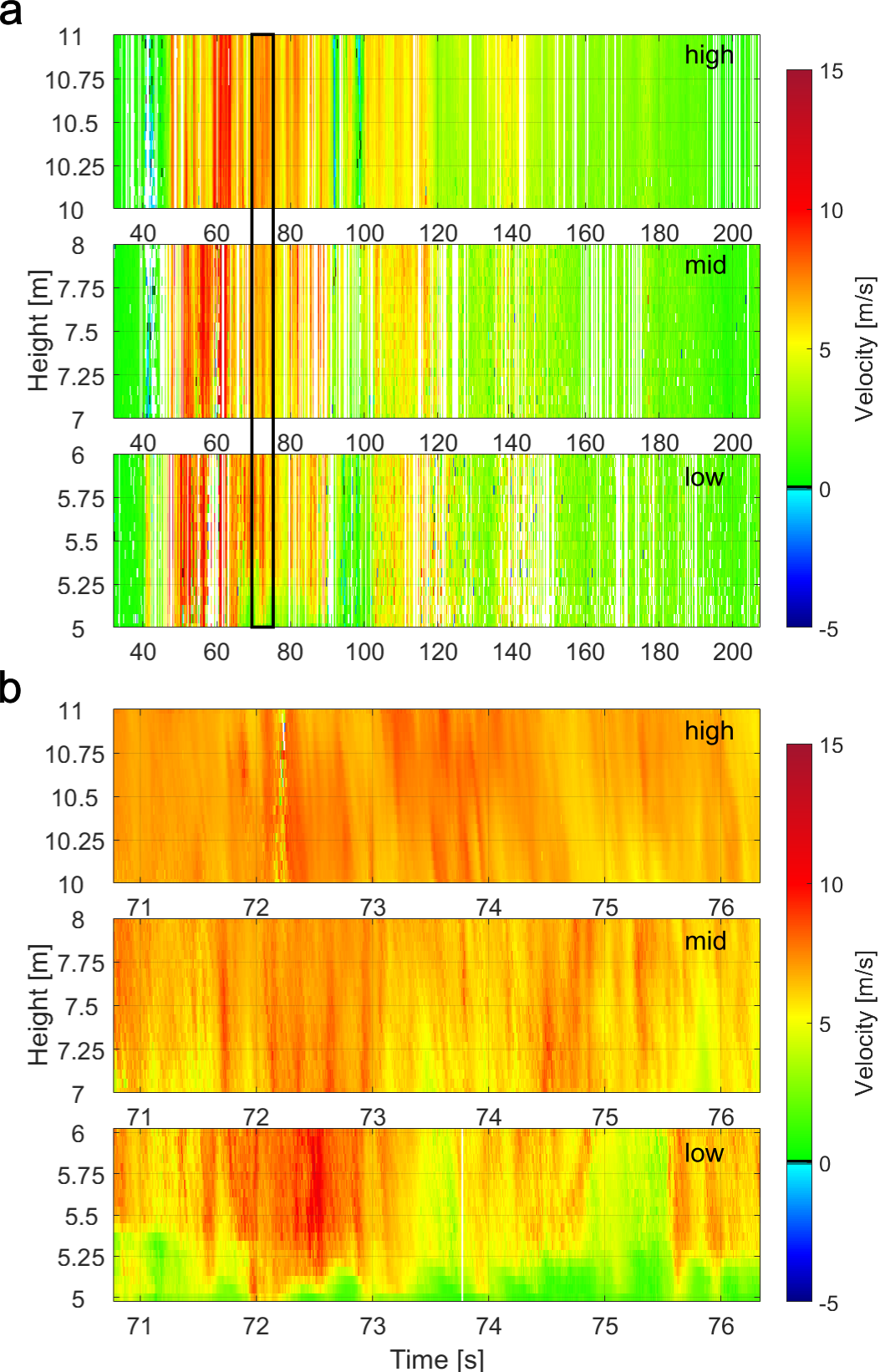}
    \caption{Space–time heatmaps of horizontal velocity obtained from PIV measurements using the high-speed camera array. a) The full velocity field for the entire high speed camera measurements at three elevation ranges above ground: high (11–10~m), mid (8–7~m), and low (6–5~m). The black rectangle marks the region selected for closer inspection. b) The corresponding zoomed-in view around 71–76~s, resolving finer flow structures and vertical velocity distributions at the same three elevations. The green signal in the lower camera corresponds to measurements of the basal layer.}
    \label{fig:PIVraw}
\end{figure}
Figure \ref{fig:PIVraw} shows the horizontal velocity field obtained from the high-speed camera array using PIV. Figure \ref{fig:PIVraw}a displays the full space-time velocity heatmap across the three stacked cameras covering the entire duration of the event. This view highlights the spatial and temporal development of the flow field, with the highest velocities recorded between 48 s and 63 s. Figure \ref{fig:PIVraw}b zooms into a 5 s interval (71–76 s), where fine-scale vertical structures and gradients become visible.

\begin{figure}[ht!]
    \centering
    \includegraphics[width=0.9\linewidth]{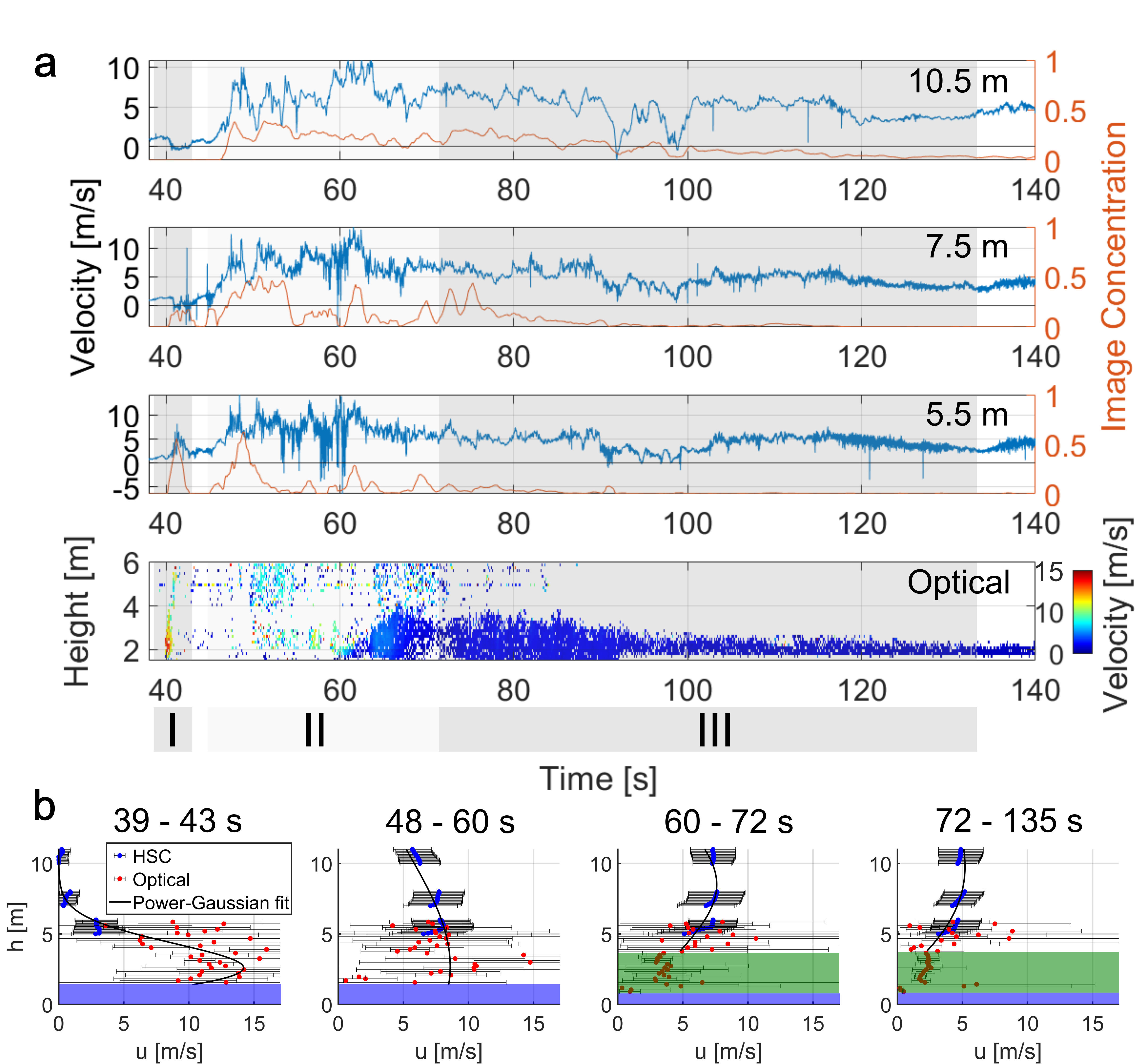}
    \caption{a) Time series of vertically averaged horizontal velocity (blue) and image concentration (red) from the three high-speed camera positions. These are labeled according to the height of the three cameras line of sight (10.5~m, 7.5~m and 5.5~m, respectively). A fourth panel shows velocities from the optical sensors. Shaded areas and the underlying bar mark the three flow regions  (I-III) as defined in Section \ref{sec:flowRegions}. b) Averaged velocity profiles for the three flow regions (I-III). Blue points denote measurements from the high-speed cameras (HSC) and red points from optical sensors. The black line is the power–Gaussian fit, with fit parameters listed in Table \ref{tab:fit}. Light-blue shading marks the height of the un-entrained snow surface, acting as the sliding bed of the flow, and the green shaded area marks the extension of the basal dense layer.}
    \label{fig:timeSeries}
\end{figure}
To condense the space–time data into more interpretable signals, velocities were vertically averaged for each camera position. Figure \ref{fig:timeSeries}a shows the resulting time series of velocities (blue curves) and image concentrations (red curves), alongside the optical sensor data. These time series capture the temporal evolution of the avalanche, including the initial onset, the fully developed airborne flow, and its subsequent decay.

Representative vertical profiles (Fig. \ref{fig:timeSeries}b) derived from three distinct flow regions are introduced in Sec. \ref{sec:vertProf}, after the regions are defined in Sec. \ref{sec:flowRegions}.

Together, these measurements provide a continuous, vertically resolved view of the avalanche across time and height, forming the basis for the analysis in the following sections.

\subsection{Definition of Flow Regions}
\label{sec:flowRegions}
\begin{figure}[ht!]
    \centering
    \includegraphics[width=0.8\linewidth]{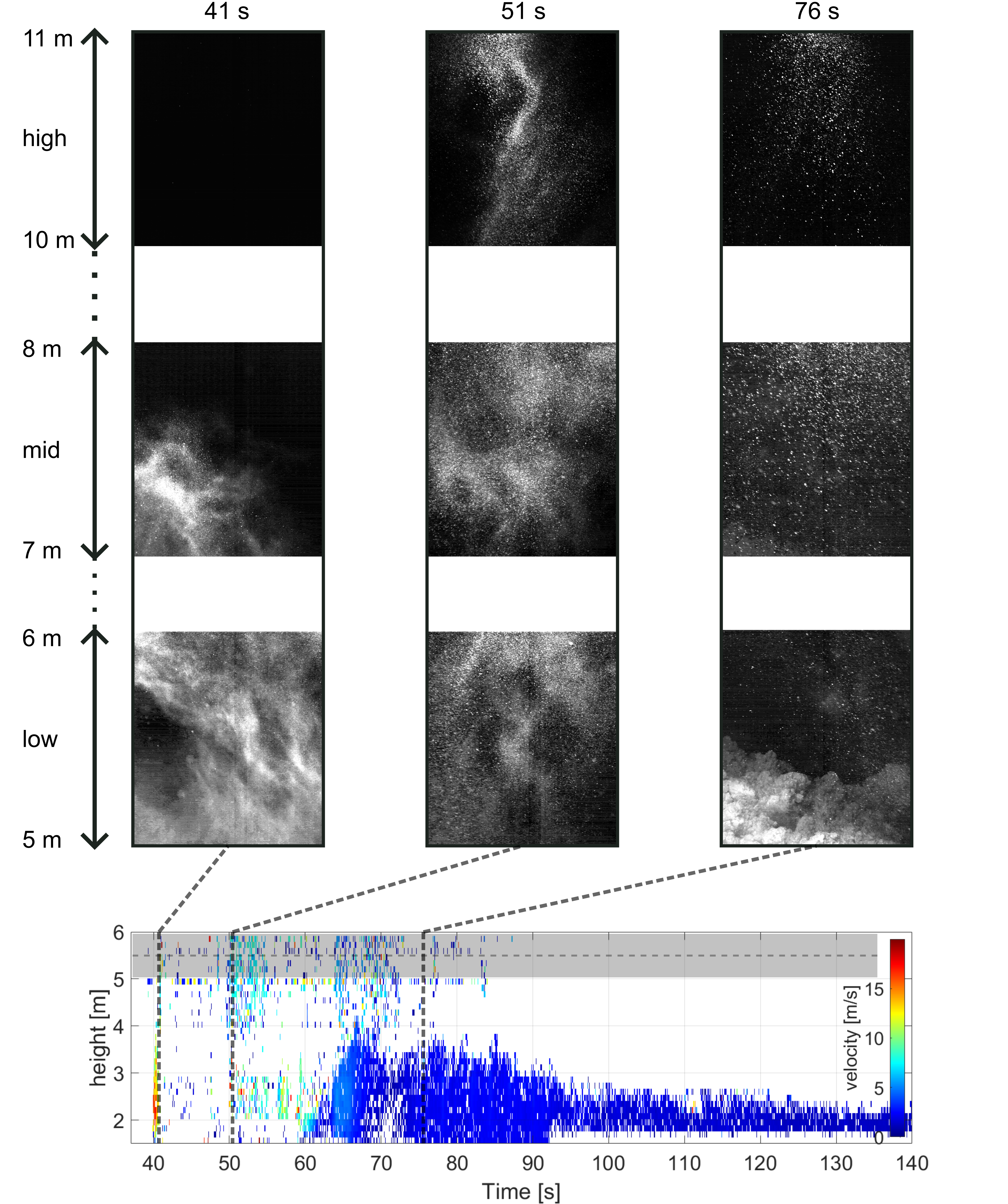}
    \caption{Selected frames from the high-speed camera array at 41~s, 51~s, and 66~s, illustrating the key flow phases: early surge, main suspension, and stabilized wake, captured at three heights (high: 11–10~m; mid: 8–7~m; low: 6–5~m). The lower panel shows velocity measurements from the optical sensors for orientation. The gray shaded box indicates the spatial overlap between the camera and optical sensor measurements and dashed lines visually link between the optical-sensor velocities with the corresponding images.}
    \label{fig:singleImages}
\end{figure}

Single-frame images from the high-speed camera array at 41~s, 51~s, and 66~s (Fig. \ref{fig:singleImages}) provide direct visual evidence of the evolving airborne flow structure at the pylon, complementing the panoramic view in Fig. \ref{fig:rawStitched} and the velocity analysis in Sec. \ref{sec:highSpeed}. The images reveal three distinct regions of the airborne flow that align with the time-series signals shown in Fig. \ref{fig:timeSeries}a, prompting the segmentation of the avalanche into the early surge (Region I), main suspension (Region II), and a stabilized wake (Region III). This classification provides the framework for subsequent analysis of velocity profiles, turbulence scales, and flow instabilities.

The early surge (Region I, 39 - 43 s), is confined to the lower two cameras, revealing a highly inhomogeneous particle distribution in the airborne layers (Fig. \ref{fig:singleImages}, 41 s). Distinct clusters of varying sizes coexist with clear voids. Some clusters appear brighter than surrounding areas and noticeably brighter than clusters observed in later phases, suggesting locally higher particle concentrations. This early surge corresponds to the first spike in the time series of velocity and image concentration (Fig. \ref{fig:timeSeries}a), observed primarily in the lower camera positions and optical sensors. The flow appears as a fast, dilute airborne front, without significant involvement of the dense basal layer, and is short-lived relative to the main body.

The main suspension (Region II, 48 - 72~s), extends across all three camera fields. Particle clusters persist but are less pronounced than during the early surge (Fig. \ref{fig:singleImages}, 51~s). This phase is characterized by higher mean velocities of around 7 m s$^{-1}$ at all three elevations, and strong fluctuations across all measured heights, while image concentration reached its overall maximum. Particle concentrations remain sufficient to generate detectable signals in the optical sensors.

In the stabilized wake (Region III, 72 - 135~s), the spatial distribution of the suspended snow particles is more dilute and homogeneous, without visible clusters of airborne material (Fig. \ref{fig:singleImages}, 76 s). The dense basal layer is visible in the lowest camera, with a sharply defined upper boundary separating it from the overlying dilute suspension. This phase corresponds to a continued decline in both velocity and image concentration, and the near disappearance of any significant signals in the optical sensor data, marking the transition to a low-energy, settling-dominated regime.

This segmentation into Regions I–III provides the basis for extracting averaged velocity profiles, which are introduced in Sec. \ref{sec:vertProf}.

\subsection{Vertical Profiles and Power-Gaussian Fitting}
\label{sec:vertProf}
To further characterize the internal flow structure, we computed region-averaged vertical velocity profiles for the three flow regions defined in Sec. \ref{sec:flowRegions} (Fig. \ref{fig:timeSeries}b). The profiles combine PIV-derived velocities from the high-speed cameras (blue points) with basal measurements from the optical sensors (red points), thereby covering the entire vertical extent of the flow accessible to our instrumentation.

The averaged velocity profiles were fitted using a modified power–Gaussian function following \citeA{brosch_destructiveness_2021}. In its original form, the model combines a power-law description for the wall region with a Gaussian relation for the jet region. Here, we adopt the extension introduced by \citeA{uhle_effects_2024}, which incorporates a basal Navier–Maxwell slip condition. This modification improves the representation of near-bed velocities while retaining the general applicability of the power–Gaussian model, which has been shown to capture vertical velocity distributions in particle-laden gravity currents \cite{cantero-chinchilla_hydrodynamic_2015, altinakar_flow_1996}. The function is expressed as: 
\begin{equation}
    u(z)=U_m\left( \frac{z + n}{h_m} \right)^{\xi}exp\left( - \left(\frac{z+n-h_m}{h_m \chi} \right)^2 -\xi \left( \frac{z+n}{h_m}-1\right)\right)
    \label{eq:fitBroschSlip}
\end{equation}
where $z$ represents the height above the avalanche bed (sliding surface), $U_m$ is the maximum velocity occurring at height $h_m$, $\xi$ and $\chi$ are exponents controlling the wall-layer and jet-region shapes, respectively and $n$ is the Navier slip length, the distance below the bed surface where the extrapolated profile would reach zero velocity. The corresponding fit parameters are listed in Tab. \ref{tab:fit}.

 \begin{table}
     \centering
     \begin{tabular}{|c|c|c|c|c|c|c|}\hline
          Region&  Time (s)&  $U_m$ (m s$^{-1}$)&  $h_m$ (m)&  $\xi$&  $\chi$& $n$ (m)\\\hline
          I&  39-43&  14.2&  2.4&  0.8&  7.4& 1.2\\\hline
 II& 48-60& 8.6& 2.4& 0.01& 28.8&1.0\\\hline
          II&  60-72&  7.6&  6.1&  0.01&  42.7& 1.4\\\hline
          III&  72-135&  5.2&  8.8&  1.0&  123.2& 2.8\\ \hline
     \end{tabular}
     \caption{Fitting parameters for the power-Gaussian velocity profile applied to three representative vertical profiles from avalanche \#~20243024 (Fig. \ref{fig:timeSeries}b). $U_m$ denotes maximum velocity, $h_m$ the corresponding height, $\chi$ and $\xi$, the jet and wall exponents, respectively and $n$ the Navier slip length.}
     \label{tab:fit}
 \end{table}

The early surge (Region I) exhibits a distinct velocity maximum of $U_m$=14.2~m s$^{-1}$ at a low height of $h_m$=2.4~m, followed by a rapid decay above the peak. This indicates that momentum is concentrated near the base, consistent with the shallow, energetic surge visible in the images (Figs. \ref{fig:rawStitched}, \ref{fig:singleImages}) and in the time series (Fig. \ref{fig:timeSeries}a). Together, these observations confirm that the flow height does not exceed the mid-camera level at 7.5~m. Near-base slip velocities are large, on the order of 10~m s$^{-1}$, while the signals from the uppermost camera reflect wind-transported snow particles rather than the avalanche body.

In the main suspension (Region II), velocity measurements extend across the entire camera range, capturing the lower portion of the gravity-current-like profile while the avalanche height exceeds the measurement limit (11~m). For the presentation of the velocity profiles, Region II is further distinguished into an early phase, where no dense basal layer is observed, and a later phase, where the basal layer is present. In the early phase of Region II, the maximum velocity is $U_m=8.6$ m s$^{-1}$ and occurs at $h_m=2.4$ m. Slip velocities are about 8 m s$^{-1}$. Compared to Region I, the maximum velocity is reduced while the peak remains near the base. In the later phase of Region II with the presence of the dense basal layer, the maximum velocity is $U_m=7.6$ m s$^{-1}$ and occurs at $h_m=6.1$ m, indicating that the velocity maximum shifts upward as the suspension cloud expands vertically. Slip velocities decrease further to about 5 m s$^{-1}$.

During the stabilized wake (Region III), the maximum velocity decreases further ($U_m$=5.2~m~s$^{-1}$) and is located at a height of $h_m$=12.6~m, reflecting a broader vertical distribution of momentum. Slip velocities are moderate (2--3~m~s$^{-1}$), while the images show the presence of the dense basal layer. These features indicate a transition toward a settling-dominated phase.

Overall, the progressive changes in maximum velocity, its height within the flow, and the magnitude of near-bed slip velocities quantify the transition from a compact surge (Region I), to a vertically expanded suspension (Region II), and finally to a diffuse, settling cloud with basal flow development (Region III). The modified power–Gaussian fitting provides a concise mathematical description of the observed profiles and demonstrates its potential as a diagnostic tool for identifying regime shifts in avalanche dynamics.

\subsection{Integral Scales of Turbulence}
To characterize the turbulence dynamics within each flow regime, we analyze the the integral length scale $L_{int}$, which represents the size of the energy-containing turbulent eddies in the flow. This quantity is derived from the integral time scale $T_{int}$, defined as the timescale over which velocity fluctuations $u'(t) = u(t) - \bar{u}$ remain correlated.

The integral time scale is obtained from the autocorrelation function  $R(\tau)$ of the horizontal PIV velocity fluctuations: 

\begin{equation}
    R(\tau)=\frac{\langle u'(t)u'(t+\tau)\rangle}{\langle u'(t)^2\rangle}
    \label{eq:autocorr}
\end{equation}

where $\tau$ is the time lag between two velocity measurements and $\langle\cdot \rangle$ denotes a time average. Following classical turbulence theory \cite{pope_turbulent_2001, monin_statistical_2013}, the integral time scale $T_{int}$ is defined as the value of $\tau$ for which $R(\tau)$ decays below $1/e$, assuming an approximately exponential decay of the autocorrelation.

Because avalanche flows are strongly non-stationary, the analysis is performed using a sliding time window of 20~s. This length was chosen conservatively to capture possible large integral scales and subsequent analysis confirmed that it exceeded the observed values. Having a window several times larger than the largest integral scale ensures robust autocorrelation estimates.

The integral length scale is then obtained by converting the temporal correlation to a spatial scale:

\begin{equation}
    L_{int}=\bar{u}\cdot T_{int}
    \label{eq:Lint}
\end{equation}

where $\bar{u}$ is the mean horizontal velocity within the corresponding time window. This conversion assumes that turbulent structures are advected at the local mean flow velocity, following a Taylor frozen turbulence hypothesis \cite{taylor_spectrum_1938}. While this assumption is strictly valid when fluctuations are small compared to the mean, and therefore may be less accurate in Region II where turbulence intensity is stronger, the main trends of $L_{int}$ are expected to remain representative of the turbulence scales. 
\begin{figure}[ht!]
    \centering
    \includegraphics[width=1.0\linewidth]{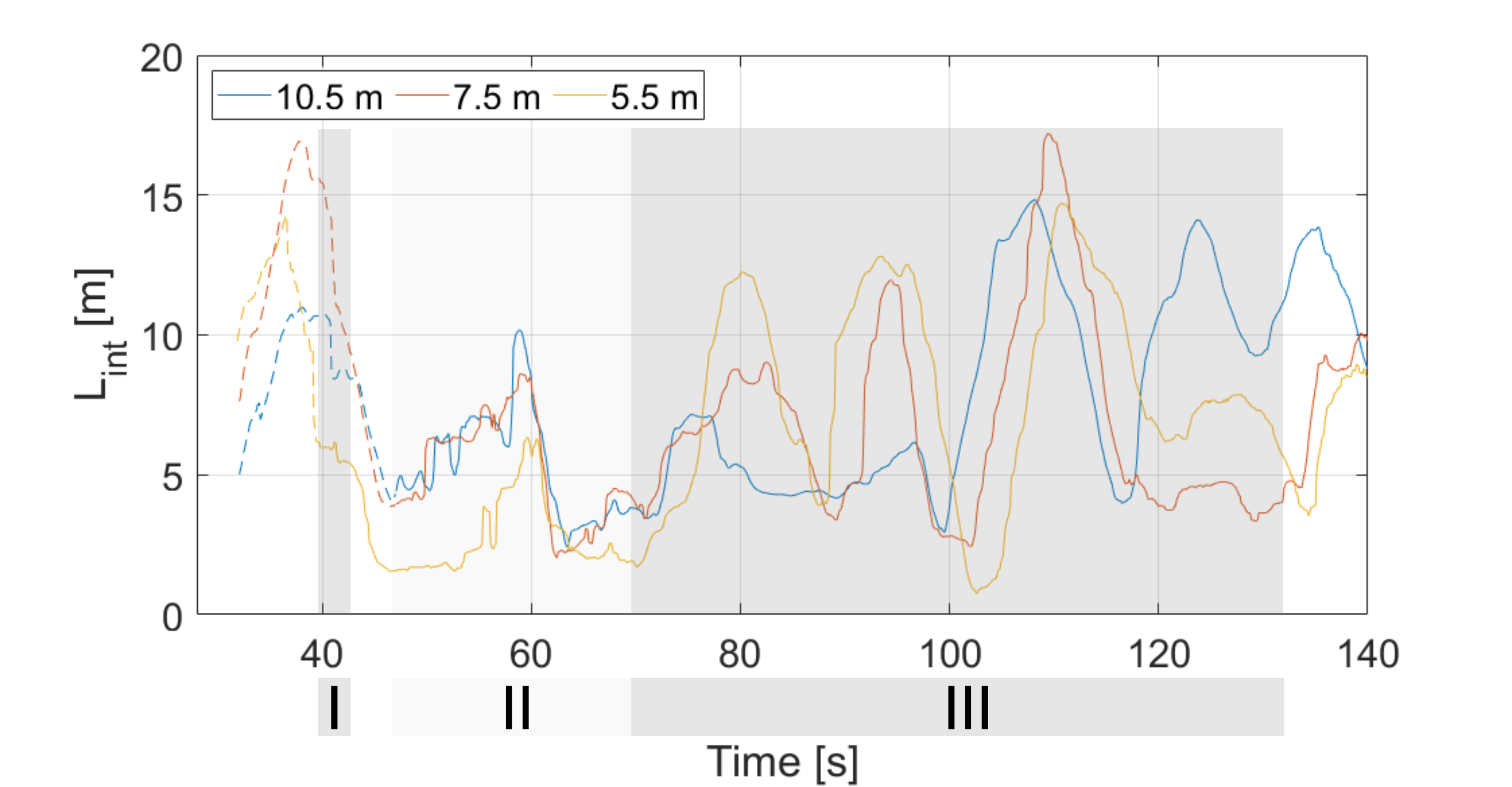}
    \caption{Temporal evolution of the integral length scale ($L_{int}$) at three heights above ground: 10.5~m (blue), 7.5~m (orange), and 5.5~m (yellow). $L_{int}$ characterizes the size of the largest turbulent eddies in the avalanche flow. Dashed lines indicate positions outside the active flow. Pre-avalanche measurements reflect ambient turbulence caused by wind and snowfall. The shaded bar below the plot marks the three flow regions (I–III).}
    \label{fig:intScale}
\end{figure}
The temporal evolution of the integral length scale ($L_{int}$) at three heights (10.5~m, 7.5~m, and 5.5~m) is shown in Fig. \ref{fig:intScale}. Prior to the avalanche, $L_{int}$ reaches values up to 17~m, reflecting ambient turbulence. This provides a reference for separating avalanche-generated turbulence from background effects.

In the early surge (Region I), only the lower camera was fully within the flow, while the mid and top cameras were either at the flow interface or outside the avalanche (dashed curves in Fig. \ref{fig:intScale}). $L_{int}$ ranges between 5--6~m at the lower position, roughly twice the height of the velocity maximum ($h_m$ = 2.4~m) from the Power-Gaussian fit and is comparable to the estimated avalanche height of 6--6.5 m.

During the main suspension phase (Region II), all camera heights were fully immersed in the flow, indicating an avalanche thickness exceeding the uppermost measurement (11~m). The integral scales fluctuate strongly, reaching maxima of 6–10~m around 60~s for the upper two cameras. At the lower position, $L_{int}$ is roughly half of the value at the upper positions, reflecting eddies confined below the velocity maximum ($h_m$ = 7.5~m). At higher positions, $L_{int}$ is larger, extending above $h_m$ but remaining smaller than the total avalanche height. This highlights the presence of two distinct turbulence scales within the PSA, associated to the shear within the dense layer below and with the ambient air above, respectively.

For the stabilized wake (Region III), the integral scales oscillate widely. The magnitudes are similar to pre-avalanche values, suggesting that in the wake of the avalanche, the observed turbulence was shaped by environmental forces (wind, ambient snow motion) rather than by intrinsic avalanche dynamics.

\subsection{Wavelet Analysis of Velocity Fluctuations}
To complement the integral scale analysis and provide a time-resolved view of turbulent fluctuations, we applied a continuous wavelet transform (CWT) to the velocity time series. The CWT decomposes the signal into contributions across different scales and times, revealing dominant fluctuation periods and their temporal evolution \cite{farge_wavelet_1992}. Among the possible wavelet functions, the complex Morlet wavelet is used as it balances time and frequency localization, making it well suited for detecting features in turbulent flows \cite{felix_combined_2005, torrence_practical_1998}. While the integral scale quantifies the dominant size of turbulent structures, wavelet analysis captures how turbulent energy is distributed over scales and how this distribution evolves with flow regions and measurement heights.  
\begin{figure}[ht!]
    \centering
    \includegraphics[width=1.0\linewidth]{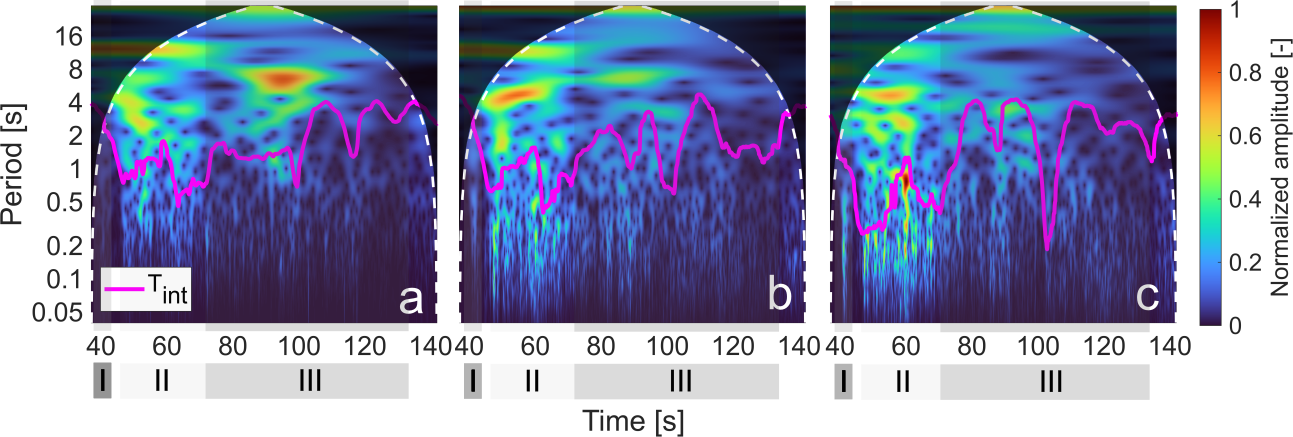}
    \caption{Wavelet scalograms of velocity fluctuations at three vertical positions: a) 10.5~m, b) 7.5~m, and c) 5.5~m. The color scale represents the wavelet power spectrum, proportional to the squared amplitude of velocity-fluctuation. The cone of influence (dashed white line) marks regions where edge effects due to the finite duration of the signal may affect the results (dark shaded). The shaded bar below the plots marks the three flow regions (I–III) defined in Fig. \ref{fig:timeSeries}. The pink curve represents the integral time scale ($T_{int}$) at each height, enabling direct comparison between the dominant turbulent structures and the integral-scale estimate.}
    \label{fig:wavelet}
\end{figure}
The scalograms in Figure \ref{fig:wavelet} illustrate the energy content associated with different time scales as a function of time. They reveal signatures of multi-scale turbulence, visible as vertically elongated patches where energy spans a wide range of fluctuation periods. These patterns indicate localized turbulent events consistent with a cascade originating at the integral time scale (pink curve). In contrast, horizontally elongated patterns correspond to more coherent, long-period fluctuations, which persist over extended time intervals.

In the early surge (Region I), the fluctuating energy is concentrated in the lowest scalogram (Fig. \ref{fig:wavelet}c), dominated by short-period motions (\(< 0.5\,\text{s}\)), all lying below the integral time scale. In contrast, in this region the mid and high positions (Fig. \ref{fig:wavelet}a, b) show little to no energy within the cone of influence, confirming that these cameras were outside or only marginally intersecting the active flow during this early phase.

As the flow enters the main suspension (Region II), all three scalograms (Fig. \ref{fig:wavelet}a–c) exhibit strong energy across multiple time scales. Near the base (lowest scalogram, Fig. \ref{fig:wavelet}c), short-period fluctuations (1~s) are consistent with the integral time scale (pink curve). Longer-period motions of 2–6~s, observed at all heights but particularly in the mid and high positions (Fig. \ref{fig:wavelet}a,b), exceed the integral time scale.

By the stabilized wake (Region III), turbulence decreases across all heights, and large-scale fluctuations (8~s) dominate at the highest position, weaker at mid height, and minimal near the base. Toward the end of this phase, dominant periods at mid and low positions converge with the integral time scale (2–3~s), signaling a shift toward smaller, turbulence-driven motions.

The CWT results demonstrate that short-period fluctuations dominate near the flow base. Coherent, longer-period motions become evident only in Region II and persist into the settling phase (Region III), particularly at mid and high elevations. Their presence at time scales above the integral time scale suggests organized motions over scales exceeding those associated with turbulent fluctuations. In earlier observations (Fig. \ref{fig:rawStitched}) we identified flow structures that visually resemble shear-driven instabilities, such as Kelvin-Helmholtz billows. Together, these combined observations motivate a more detailed investigation of flow stability, presented in the next section.

 \subsection{Instability Analysis}
To assess whether the measured flow conditions could support shear-driven instabilities, we follow \citeA{negretti_linear_2008} and \citeA{kostaschuk_causes_2018} and evaluate both the bulk Richardson number $J$ and the nondimensional wavenumber $\alpha_r$. We compute values of $J$ and $\alpha_r$ using horizontal velocity profiles averaged over 0.5~s windows. For each profile, $\Delta U$ is defined as the difference between the maximum and minimum values across the profile, while $(dU/dz)_{max}$ is the largest velocity gradient. The bulk Richardson number, which determines whether a sheared interface between two fluids is stable \cite{ellison_turbulent_1959}, is given by: 

\begin{equation}
    J=\frac{\Delta \rho g \delta U}{\rho_0 \Delta U ^2}
\end{equation}
where $\Delta \rho = \rho_s - \rho_f$ is the density difference between the suspension layer $\rho_s$ and the ambient air $\rho_f$,  $\rho_0 = (\rho_s + \rho_f)/2$ is the mean density, and $g$ is the gravitational acceleration. The air density $\rho_f$ is estimated from temperature measurements. The suspension-layer density is approximated as $\rho_s=2.2$ kg m$^{-3}$ obtained from a pressure sensor mounted 11~m above ground on the pylon, by matching the dynamic pressure $p(t)=\frac{1}{2}\rho_s u(t)^2$ to velocities derived from high-speed camera data (see \ref{sec:susDens}). For a discussion on how the dynamic pressure is related to the mean suspension density (excluding anomalously high peaks due to individual particle impacts), see \citeA{uhle_turbulent_2024}.

The shear layer thickness is approximated as $\delta U=\frac{\Delta U}{dU/dz}$.

The nondimensional wavenumber is defined as:

\begin{equation}
    \alpha_r = \frac{2\pi}{\lambda}\frac{\Delta U}{(dU/dz)_{max}}.
    \label{eq:wavenumber}
\end{equation}

where $\lambda = \tau_m\bar{u}$ is the wavelength and $\tau_m$ is the period of the low-frequency oscillations identified in the wavelet scalogram (Fig. \ref{fig:wavelet}). 
 
\begin{figure}[ht!]
    \centering
    \includegraphics[width=0.8\linewidth]{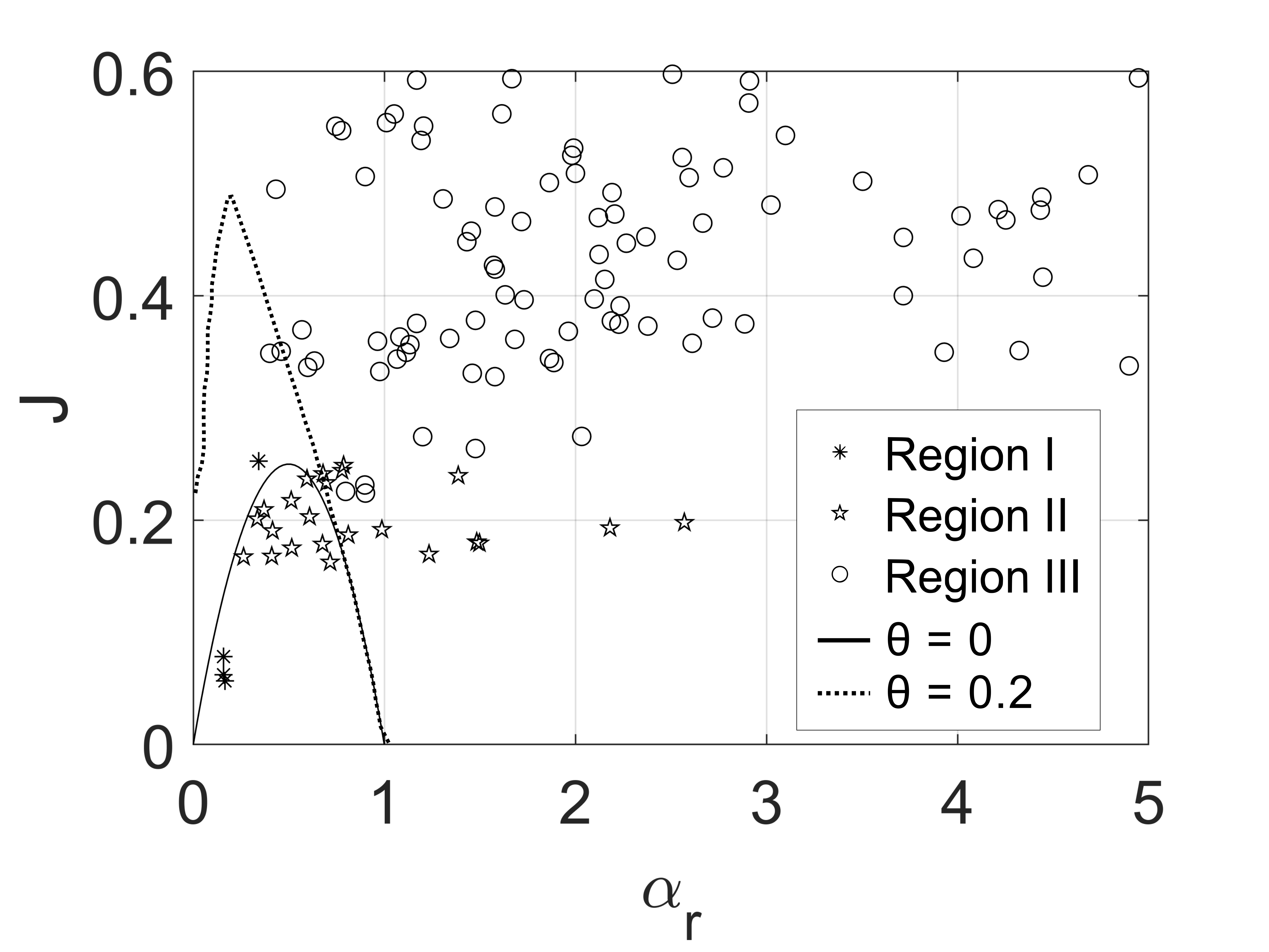}
    \caption{Stability analysis based on velocity measurements. Symbols show calculated values of bulk Richardson number $J$ and nondimensional wavenumber $\alpha_r$ from velocity profiles averaged every 0.5~s, grouped by region (I–III). Neutral stability curves  from \citeA{negretti_linear_2008} are shown for slope angles $\theta$=0~rad (solid line) and $\theta$~=0.2~rad (dashed line). Points below the curves fall within the unstable regime.}
    \label{fig:instability}
\end{figure}
Figure \ref{fig:instability} compares the resulting values of $J$ and $\alpha_r$ to the neutral stability curves derived by \citeA{negretti_linear_2008}. The solid curve corresponds to a slope angle of $\theta$=0~rad, reducing to the analytical Taylor–Goldstein solution $J = \alpha_r(1-\alpha_r)$ \cite{holmboe_behavior_1962, scotti_experiment_1972, eriksen_observations_1982, kundu_fluid_2024}, while the dashed curve corresponds to $\theta$= 0.2~rad, obtained numerically. These curves divide the $J-\alpha_r$ plane into stable regimes above the curves, where perturbations decay, and unstable regimes below, where perturbations grow.

Although the slope in the measurement area exceeds 0.2~rad (11.5°), the terrain around the pylon is nearly flat making the 0.2~rad curve a useful reference. The data points show a systematic evolution across flow regions: Region I lies within or close to the theoretically unstable regime, Region II lies near the stability boundary, and Region III is mostly stable. This progression is consistent with increasing stratification and the transition toward a settling-dominated flow.

It should be noted that the velocity profiles used to compute $J$ and $\alpha_r$ are derived from flow fields measured during the evolving avalanche. Therefore, the stability analysis represents an a posteriori diagnostic of the instantaneous flow conditions, rather than a prediction of the initial onset of the instability. The results should thus be interpreted as indicating that the observed flow states are consistent with conditions that allow shear-driven instabilities to persist or develop, rather than demonstrating their triggering.
 
\subsection{Flow Regimes, Turbulence and Particle Dynamics}
\label{sec:flowRegTurPar}
This section uses high-resolution velocity and particle measurements to estimate key dimensional and nondimensional parameters, providing a quantitative characterization of the multiphase flow regimes, turbulence, and particle dynamics in the avalanche. Such measurements, previously inaccessible in conventional field observations, have only rarely been obtained in studies of TCs, PDCs, and PSAs \cite{meiburg_turbidity_2012, brosch_destructiveness_2021}.
\begin{figure}[ht!]
    \centering
    \includegraphics[width=1.0\linewidth]{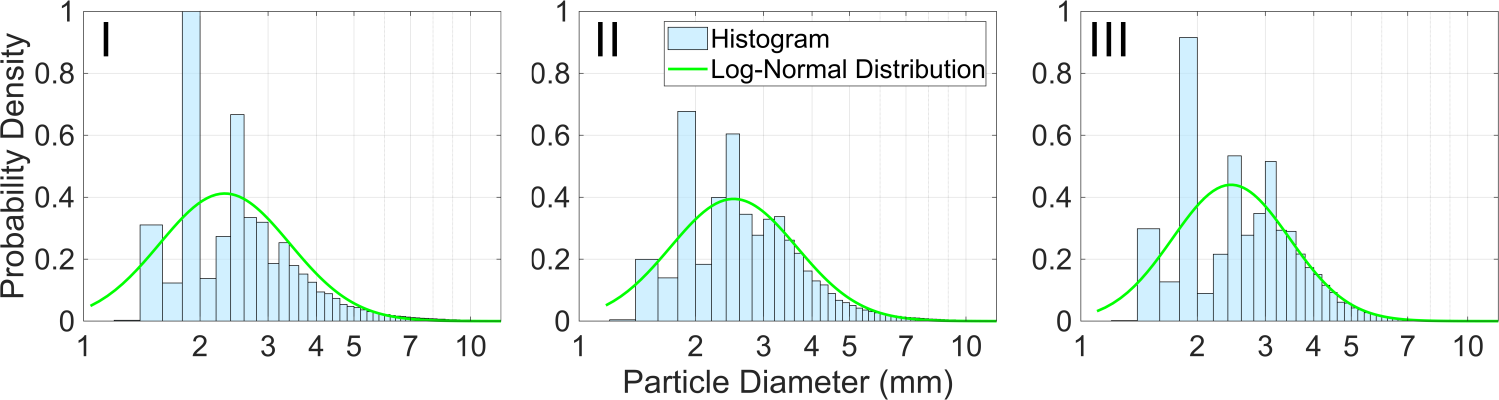}
    \caption{Probability density functions of particle diameters for Regions I–III of avalanche \#~20243024. Blue bars show normalized histograms of detected particle sizes, and green curves represent fitted log-normal distributions. Vertical axis is normalized to unit area; horizontal axis is on a logarithmic scale. Geometric mean diameters for each region are given in Tab. \ref{tab:dimensional}.}
    \label{fig:PSDregions}
\end{figure}

We begin with the characterization of the snow particle sizes. Figure \ref{fig:PSDregions} shows the particle size distributions for Regions I–III of avalanche \# 20243024. All distributions follow approximately log-normal shapes, with geometric mean diameters between 2.7~mm and 2.9~mm. The mean values, reported in Tab. \ref{tab:dimensional} along with other key dimensional parameters, serve as input for estimating particle aerodynamic response times and settling velocities\textit{. }As mentioned in section \ref{sec:imagBased}, in Region I the high concentration and high velocity complicates the particle identification and sizing. The similar size distributions across the entire recording, however, suggest that the estimated ranges are robust.

The snow particle density is taken as $\rho_p = 300$~kg~m$^{-3}$ \cite{mcclung_characteristics_1985, gauer_snow_2008}. 
Air properties including density $\rho_f$ and kinematic viscosity $\nu$ are estimated based on air temperature measurements (ambient temperature of $T_a = -3$°C).  

The particle aerodynamic response time is calculated as:

\begin{equation}
    \tau_p = \frac{\rho_p d_p^2}{18 \mu (1+0.15 Re_p^{0.687})}, 
    \label{eq:taup}
\end{equation}

using the Schiller–Neumann correction for the finite particle Reynolds number $Re_p = \frac{V_td_p}{\nu}$ \cite{clift_bubbles_2005}, solved via a simple iterative approach \cite{tanaka_sub-kolmogorov_2010}. The still-air settling velocity of the particles is estimated as:

\begin{equation}
    V_t = \tau_p g
\end{equation}
\cite{brandt_particle-laden_2022}.
The root-mean-square velocity fluctuations calculated from the PIV velocity time series $u_{RMS}$ (m/s) and the integral length scale $L_{int}$ are used to estimate the turbulent dissipation rate:

\begin{equation}
    \epsilon = u_{RMS}^3/L_{int}
\end{equation}

\cite{tennekes_first_1972}. In doing this, we assume that the large-scale properties of the turbulence are captured by the particle motion, which is supported by evaluating the Stokes number, below. The Kolmogorov time scale of the dissipative eddies is calculated as:
 \begin{equation}
     \tau_{\eta} = (\nu / \epsilon)^{1/2}
 \end{equation}

\begin{table}
    \centering
    \begin{tabular}{|c|c|c|c|c|}\hline
         &  &  Region I&  Region II& Region II\\\hline
         Particle diameter&  $d_p$ (mm)&  2.7&  2.9& 2.8\\\hline
         Particle response time&  $\tau_p$ (s)&  0.42&  0.45& 0.44\\\hline
         Particle setting velocity&  $V_t$ (m s$^{-1}$)&  4.19&  4.44& 4.32\\\hline
         Mean velocity&  $\bar{u}$ (m s$^{-1}$)&  3.3 (11.6)&  7.1& 4.6\\\hline
         Root-mean-square velocity&  $u_{RMS}$ (m~s$^{-1}$)&  2.5&  1.7& 0.9\\\hline
         Dissipation rate&  $\epsilon$ (m$^2$s$^{-3}$)&  2.7&  1.1& 0.1\\\hline
         Kolmogorov timescale&  $\tau_{\eta}$ (s)&  3.6 $\cdot$10$^{-5}$ &  1.5 $\cdot$10$^{-5}$& 1.3 $\cdot$10$^{-6}$\\\hline
         Mean integral time scale&  $\bar{T}_{int}$ (s)&  1.12&  0.93& 2.06\\\hline
 Mean integral length scale& $\bar{L}_{int}$ (m)& 5.7& 4.4&7.7\\ \hline
    \end{tabular}
    \caption{Dimensional flow and particle parameters for the three identified regions (I–III) of avalanche \# 20243024. These values include particle sizes, settling velocities, velocity fluctuations, and turbulence scales, providing the physical basis for estimating particle response and turbulent coupling in the airborne layers. The mean velocity in parentheses is derived from the optical sensors. Parameters that do not vary across regions are: particle density $\rho_p$ (300~kg~m$^{-3}$), air density $\rho_f$ (1.27~kg~m$^{-3}$), air kinematic viscosity  (1.33$\cdot$10$^{-5}$~m$^2$~s$^{-1}$ ), and suspension density $\rho_s$ (2.2~kg~m$^{-3}$).}
    \label{tab:dimensional}
\end{table}
Furthermore, in Table \ref{tab:dimensionless} we report a suite of non-dimensional parameters in the three identified regions of the avalanche, using values averaged over their corresponding time spans. 

To assess how particles interact with turbulence, we compare particle and flow time scales. The Stokes number based on the integral and Kolmogorov time scales are defined as:

\begin{equation}
    St_{int} = \tau_p /T_{int},
\end{equation}
and
\begin{equation}
    St_{\eta} = \tau_p/\tau_{\eta},
\end{equation}

respectively. In all regions, $ St_{int}<1$, implying that snow particles respond to the energetic eddies of the turbulence. In contrast, the relatively large $ St_{\eta}\gg 1$ indicates that the particles behave nearly ballistically with respect to fine-scale turbulent fluctuations \cite{petersen_experimental_2019, brandt_particle-laden_2022}.

The Reynolds number,
\begin{equation}
    Re = \bar{u}L_{int}/\nu,
\end{equation}
quantifies the ratio of inertial to viscous forces. Across all regions, Re lies in the range of $O(10^5-10^6)$, characteristic of large-scale geophysical flows. The decrease from Region II to Region III reflects progressive decay of the strength of the flow and the intensity of the turbulence.

The Richardson number $J=O(0.1)$ in Region I indicates conditions favorable for shear-driven turbulence and instabilities, while the increase in Regions II and III reflects the strengthening of stratification as the flow decelerates. As it was argued that $J$ is the main factor determining volume growth rate of PSAs \cite{ancey_powder_2004}, the trend may also indicate decreasing snow entrainment from the bed.

The settling parameters $V_t/\bar{u}$ and $ V_t/u_{RMS}$ quantify the relative importance of gravitational settling versus mean advection and turbulent suspension \cite{balachandar_turbulent_2010}. In Regions I and II these ratios are much smaller than unity, confirming strong turbulent suspension. In Region III, however, settling becomes comparable to both the mean and fluctuating velocities, indicating that deposition processes become important. 

\begin{table}
    \centering
    \begin{tabular}{|c|c|c|c|c|}\hline
         &  &  Region I&  Region II& Region III\\\hline
         Particle Reynolds number&  $Re_p$&  850&  968& 908\\\hline
         Flow Reynolds number&  $Re$&  1.3 (4.9) $\cdot$10$^6$&  2.3 $\cdot$10$^6$& 2.4  $\cdot$10$^6$\\\hline
         Bulk Richardson number&  $J$&  0.12&  0.2& 0.8\\\hline
         \makecell{ Stokes number \\ (based on integral scale) }&  $St_{int}$&  0.4&  0.6& 0.2\\\hline
         \makecell{Stokes number \\ (based on Kolmogorov scale) }&  $St_{\eta}$&  190&  129& 37\\\hline
         \makecell{Settling parameter \\ (based on mean velocity)}&  $V_t/\bar{u}$&  1.4 (0.4)&  0.6& 0.8\\\hline
         \makecell{Settling parameter \\ (based on fluctuating velocity)}&  $V_t/u_{RMS}$&  1.7&  2.6& 4.8\\ \hline
    \end{tabular}
    \caption{Dimensionless numbers characterizing flow regimes in Regions I–III of avalanche \#~20243024. The values quantify turbulence intensity, stratification, and particle–fluid interactions, offering a comparative framework for interpreting regime transitions across the avalanche. The Reynolds number shown in parentheses is derived from the optical sensors.}
    \label{tab:dimensionless}
\end{table}

\section{Discussion}
\subsection{Avalanche Structure and Flow Regimes}
Avalanche \#~20243024 displayed an airborne structure that deviated from the classical three-layer model of \citeA{sovilla_structure_2015}, which distinguishes a dense basal layer, a transitional layer, and an upper suspension layer. In this relatively small PSA, we observed a dense basal layer with an overlying suspension, but no distinct transitional layer. This suggests that limited entrainment from the bed prevented the formation of an intermediate region, leaving the suspension dominated by air entrainment and subsequent particle settling.

Ahead of the main avalanche body, GEODAR data and high-speed camera observations captured short-lived, compact surges (Region I). These were followed by a main body in which particles remained strongly coupled to the turbulence and intermittently clustered (Region II), followed by a wake in which  particle motion was largely controlled by gravitational settling (Region III). In Regions I and II velocity fluctuations were intense and the spatial distribution of suspended snow particles was highly heterogeneous, challenging the traditional view  that the suspension layer is relatively uniform \cite{sovilla_structure_2015, issler_inferences_2020}. 
 The measured nondimensional parameters support this interpretation: the Richardson number and the settling parameter were consistent with an unstable flow in which turbulent motions maintained particles in suspension. In Region III, on the other hand, the same parameters signaled a more stably stratified flow dominated by particle settling. These transitions resembled patterns observed in laboratory density currents \cite{cantero-chinchilla_hydrodynamic_2015} and large-scale pyroclastic analogs \cite{breard_inside_2017, lube_multiphase_2020}.

Turbulence properties further highlight the evolving dynamics of the avalanche. During Region I, the integral length scales were comparable to the flow depth, reminiscent of the head of particle-laden gravity currents where depth-filling eddies promote strong mixing and entrainment of ambient fluid \cite{ellison_turbulent_1959, turner_buoyancy_1973, parker_experiments_1987, necker_high-resolution_2002}. In contrast, in Region II the energy-containing eddies generated by shear along the bed remained well below the airborne layer height, reflecting stratification-limited turbulence similar to that observed in TCs at relatively large $J$ \cite{lacaze_gravity_2024}. Larger scales detected farther from the bed indicated significant turbulence production by the shear with the ambient air above, similar to observations in PDCs \cite{brosch_destructiveness_2021}. The oscillations of the integral scales in Region III appeared driven by atmospheric turbulence, likely reflecting the influence of katabatic winds \cite{denby_use_2000, umphrey_direct_2017}. 

\subsection{Flow Instabilities}
The wavelet scalograms highlight the wide range of scales associated with the flow. Especially in Region II, high-energy fluctuation periods consistently exceeded the integral time scale, indicating the presence of coherent motions beyond the turbulent cascade. Visual evidence of billow-like structures further supports this interpretation (Fig. \ref{fig:rawStitched}). Moreover, the stability analysis suggests that these long-lived structures are consistent with shear-driven KH-type rollers in the airborne layer. As these contain a substantial fraction of the fluctuating energy, they may act as a primary driver of turbulence within the active suspension, even though the present data does not allow a direct link to the dynamics of the turbulent cascade.

Shear instabilities have long been hypothesized to influence the evolution of PSAs based on remote imagery, laboratory experiments, and modeling \cite{hopfinger_model_1977, hutter_avalanche_1996, billingham_dam_2019}, but direct confirmation under natural conditions has been lacking. Our in situ measurements represent a strong indication of their role in a full-scale setting. Similar mechanisms were shown to enhance entrainment and promote momentum exchange in TCs and PDCs \cite{meiburg_turbidity_2010, breard_inside_2017, lube_multiphase_2020}, further highlighting analogies across these different classes of natural flows, despite their distinct respective features.

\subsection{Particle Size, Inertia and Spatial Distribution}
Snow particle sizes in PSAs were rarely measured, and the reported ranges vary widely. \citeA{schaer_particle_2001} inferred particle sizes from impact pressure measurements, obtaining values from millimeters to centimeters. Those, however, supposedly reflected particles in the transition layer, which was absent in the avalanche we observed. \citeA{rastello_size_2011}, on the other hand, collected particles at the VdlS bunker and found a geometric mean diameter of 0.16 mm. This comparatively fine distribution likely reflected the sampling location, which is typically reached only by the wake region, where coarse fragments have already settled and only the smallest particles remain aloft. \citeA{ito_measurement_2017} measured sizes around 0.2 mm using an optical snow-particle counter in a small mixed avalanche. In this case, the relatively fine values might be due partly to the instrument’s upper detection limit, and partly to the short avalanche path limiting the development of energetic motions capable of suspending larger particles. Our in-situ, imaging-based measurements of airborne particles fall in the wide range of previously reported sizes, and add data points specific to the features of the investigated PSA.

For the purpose of interpreting and modeling the PSA regimes, the crucial feature of snow particles is their inertia, embodied by the aerodynamic response time based on particle size and density. While the latter is inevitably associated with large uncertainties, a first-order estimate provides important indications from the range of values of the Stokes number. As $St_{int}= O(1)$, the particles are expected to respond to the energy-containing eddies, which in turn justifies the approximation of using the statistics of their motion to evaluate the properties of the multiphase turbulent flow. This is also consistent with the strongly non-uniform spatial distribution of the airborne particles. Although the limited FOVs do not allow for a full analysis of the length scales of particle clusters, it is evident from the recordings that patches of highly concentrated material reached meter-scale extensions, alternating to similarly large voids. Such unsteadiness in mass transport likely impacts the large pressure oscillations typical of PSA \cite{sovilla_structure_2015}.

\subsection{Implications for Modeling}
Several aspects of the present observations, while influenced by the specific features of the investigated PSA, may help inform and constrain numerical and reduced-order models. The first is the profoundly different dynamics across the various phases of the flow. While the segmentation in the three proposed regions is not univocal, it is clear that strong qualitative differences exist between the early surge, the turbulent body and the settling-dominated wake. Likewise, although a modified power–Gaussian fit to the velocity profile can describe the three regions, its parameters vary substantially between those. The data also highlight a significant basal slip which varies across the regions and is expected to influence near-bed momentum transfer and snow entrainment. Realistic parameterizations should allow such changes across the evolving flow, which classic depth-averaged and box-model approaches do not account for \cite{ancey_snow_2001}.

Second, the inertial nature of the suspended snow particles cannot be ignored. Their level of inertia results in highly non-uniform concentration, and therefore the assumption of well-mixed current is not tenable. This implies that the applications of shallow water models is also questionable, although it was recently shown to reproduce certain features of PSAs (under ad hoc assumptions of entrainment rate and source of mass, \citeA{ivanova_numerical_2022}). More fundamentally, the relatively large Stokes number implies that, in a PSA such as the one we investigated, snow particles cannot be rationalized as flow tracers to which a fall speed is superposed, nor are they amenable to simplified models for weakly inertial particles \cite{cantero_turbulent_2008}. In this perspective, even Reynolds-Averaged Navier-Stokes (RANS) models would not capture the important dynamics, and Euler-Lagrange approaches such as Large-Eddy Simulations (LES) coupled with inertial particle transport \cite{meiburg_modeling_2015} may be needed, though at a large computational cost. The present data may also inform parameterizations needed in Euler-Euler simulations where the air and particle phases are treated as inter-penetrating fluids, similar to recent PDCs simulations \cite{breard_continuum_2019}.

Finally, the unsteady nature of the PSA is not only evident in its decaying strength but also in the wide oscillations of its properties. These have been recently highlighted in various gravity driven currents thanks to novel field observations. \citeA{brosch_destructiveness_2021} showed how in PDCs the dynamic pressure reflects large-scale coherent turbulent structures and internal gravity waves, producing low-frequency, high-pressure pulses. In the present PSA, pulsations arise not only from turbulence-induced fluctuations but also from flow instabilities and surges propagating ahead of and inside the main body. Because of the importance of such fluctuations in the avalanche destructive potential, advanced models shall aim to capture their magnitude and dominant frequency.

\section{Conclusion}
In this study we presented the first direct, high-resolution optical observations of particle-scale dynamics inside the airborne layers of a natural PSA. By combining GEODAR, optical sensors, and high-speed imaging, we were able to resolve vertical velocity structures, turbulence properties, particle distributions, and signatures of instabilities across the event. Segmenting the avalanche into three regions provided a clear framework for analyzing how airborne flow regimes evolve in space and time.

These observations show that PSAs are far from uniform. Turbulence and shear-driven instabilities were found to be central to sustaining the airborne suspension and driving clustering, while particle inertia produced highly heterogeneous concentration fields and ultimately a transition toward settling.
The analyzed avalanche was relatively small and partially decelerating compared to high-magnitude hazard events. Nevertheless, the identified instability and turbulence–particle coupling mechanisms are expected to persist in larger PSAs, even if their absolute intensity increases and associated model parameters differ. Together, these findings refine the conceptual model of avalanche structure and provide rare empirical benchmarks for advancing multiphase gravity current models. They also demonstrate strong analogies with turbidity and pyroclastic currents, suggesting broader relevance for particle-laden geophysical flows. 

Looking ahead, our results highlight the need for models that explicitly represent turbulence–particle interactions and evolving instabilities, rather than relying on simplified closures. Extending high-resolution measurements across avalanches of different sizes and snow conditions will be essential to capture the variability of natural flows. More broadly, the methodology introduced here opens new opportunities to advance the general physics of multi-phase gravity currents which is important for improving hazard forecasting and mitigation strategies.

\appendix
\section{Suspension Density}
\label{sec:susDens}

\begin{figure}[ht!]
    \centering
    \includegraphics[width=1\linewidth]{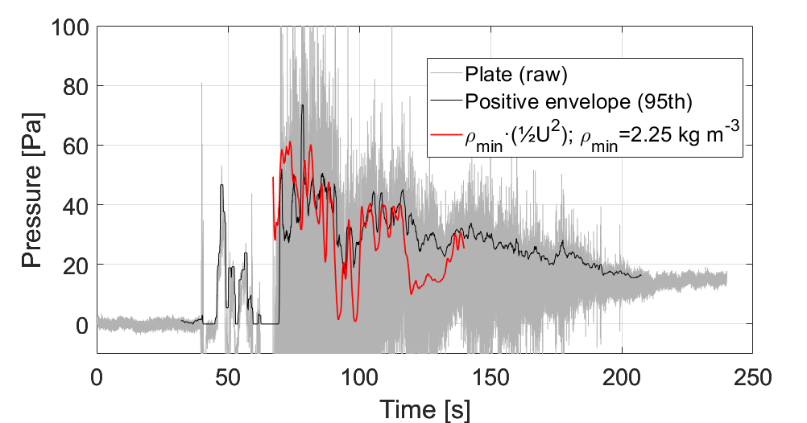}
    \caption{Time series of raw plate pressure (gray) with the 95th-percentile positive envelope (black) and the calculated dynamic pressure (red)}
    \label{fig:combi}
\end{figure}

The positive peaks of the plate signal in the post-dip interval correspond to brief, near-stagnation collisions of the air-snow mixture on the plate. Those moments are the ones that should scale with the dynamic pressure term of $1/2  \ U^2$. To isolate them, we take a positive envelope of the pressure (at 95 percentile), which captures the tops of the bursts while suppressing suction and recirculation periods, as shown in Fig. \ref{fig:combi}. 

We then relate the envelope pressure to velocity as

\begin{equation*}
    p_{env} \approx C_{eff}\rho\frac{1}{2}U(t)^2,
\end{equation*}

where $C_{eff}\in(0,1]$ accounts for departures from ideal stagnation. Setting $C_{eff}=1$ gives a conservative lower bound:

\begin{equation*}
    \rho_{min}=\frac{p_{env}}{\frac{1}{2}U^2},
\end{equation*}

The true density satisfies $\rho_{true} = \rho_{min}/C_{eff }\geq \rho_{min}$.
 
To obtain a single value, we fit the slope between the positive envelope and the dynamic pressure term. We regress

\begin{equation*}
    p_{env}(t) = \rho_{min}(\frac{1}{2}U(t)^2).
\end{equation*}

The result is illustrated in Fig.~\ref{fig:combi}, which compares the positive envelope of the measured plate pressure with the calculated dynamic pressure using $\rho_{min}=2.25$~kg~m$^{-3}$. The good agreement between the two curves confirms that this value provides a consistent lower-bound estimate of the bulk suspension density.

%



%
%

\section*{Open Research Section}
The GEODAR radar, optical velocity and pressure plate data, together with concentration and velocity data from the high-speed camera images are available at the repository ZENODO, via the following link:
https://doi.org/10.5281/zenodo.17104410

\acknowledgments
This work was supported via the Swiss National Science Foundation project 'The pulsing nature of powder snow avalanches' under Grant No. 207323. The data from the optical sensors was acquired as part of the Swiss National Science Foundation grant No. 113069  (Snow Avalanches in the 'Swiss Experiment'). 
The GEODAR data was acquired as part of the Swiss National Science Foundation grant No. 143435. We thank Pierre Huguenin and Anselm Köhler for the processing of the GEODAR data.
We gratefully acknowledge the excellent support from the Experimental Facility Group at the WSL Institute for Snow and Avalanche Research SLF as well as from the canton of Valais.

\noindent
\textbf{Declaration of interests:} The authors declare there are no conflicts of interest for this manuscript.

%
%
\bibliography{references}

%
%
%
%
%

\end{document}